\documentstyle[11pt,paspconf,psfig,epsf]{article}

\newcommand{\QrhoNote}{\footnote{A spheroidal system with
shortest-to-longest axis ratio $c/a \; (q_{\rho})$ of the density
contours has a shape E$n$, with $n$ such that $q_{\rho}=1-n/10$}}
\newcommand{\VLAnote}{\footnote{The VLA of the
National Radio Astronomy Observatory is a facility of the National
Science Foundation operated under cooperative agreement by Associated
Universities, Inc.} \ }
\newcommand{\etal}{\mbox{{\it et al. }}}
\newcommand{\qrho}{\mbox{$q_{\rho}$}}
\newcommand{\pmt}{\mbox{$\pm \;$}}
\newcommand{\rhoN}{\mbox{$\rho_0$}}
\newcommand{\Rc}{\mbox{$R_{\rm c}$}}
\newcommand{\Kz}[1]{\mbox{$K_{\rm z}{#1}$}}
\newcommand{\kms}{\mbox{${\rm km \;s}^{-1}$}}
\newcommand{\HI}{\mbox{{\rm H \footnotesize{I} }}}
\newcommand{\MoverL}[1]{\mbox{$M/{\cal L}_{{\rm #1}} \;$}}
\newcommand{\Msun}{\mbox{$M_{\odot}$}}
\newcommand{\rtp}[1]{\mbox{$^{#1}$}}
\newcommand{\MSpcsq}{\mbox{$M_{\odot}{\rm pc}^{-2}$}}

\def\edcomment#1{\iffalse\marginpar{\raggedright\sl#1\/}\else\relax\fi}
\reversemarginpar

\begin{document}

\title{The Highly Flattened Dark Matter Halo of NGC 4244}

\author{Rob P. Olling}
\affil{Columbia University, now at University of Southampton, Department of
Physics and Astronomy, Southampton SO17 1BJ, United Kingdom}

\author{To appear in "Dark and Visible Matter in Galaxies and Cosmological
Implications," eds. M. Persic and P. Salucci, A.S.P. Conference Series. 1997}

\author{}
\affil{}

\vspace*{-7mm}
\begin{abstract}

In a previous paper (Olling 1995) a method was developed to determine
the shapes of dark matter halos of spiral galaxies from an accurate
determination of the rotation curve, the flaring of the gas layer and
the velocity dispersion in the \HI.  Here I report the results for the
almost edge-on Scd galaxy NGC 4244 (Olling 1996a, 1996b). 

The observed flaring of the \HI beyond the optical disk puts significant
constraints on the shape of the dark matter halo, which are almost
independent of the stellar mass-to-light ratio.  NGC 4244's dark matter
halo is found to be highly flattened with a shortest-to-longest axis
ratio of $0.2_{-0.1}^{+0.3}$.  If the dark matter is disk-like, the data
presented in this paper imply that the vertical velocity dispersion of
the dark matter must be 10\% - 30\% larger than the measured tangential
dispersion in the \HI.

\end{abstract}

\vspace*{-7mm}
\section{Introduction}

Although rotation curves of spiral galaxies have been used as evidence
for the presence of dark matter (DM), little is known about the nature,
extent and actual distribution of the DM in individual galaxies (e.g.,
van Albada \etal 1985; Lake \& Feinswog 1989).  As measurements of the
equatorial rotation curve probe the potential in only one direction,
they provide no information about the shape of the DM halos. 
 
Several methods have been used to determine the shapes of dark matter
halos.  Analyzing the warping behavior of \HI disks, Hofner \& Sparke
(1994) conclude that only one (NGC2903) of the five systems studied
requires a DM halo as flattened as E4\QrhoNote.  On the other hand, in
studies of polar ring galaxies (Sackett \& Sparke 1990; Sackett \etal
1994; Sackett \& Pogge 1995) substantially flattened DM halos are found
(E6-E7 for NGC 4650A, E5 for A0136-0801).  The shape of the dark halo of
the Milky Way has been estimated (E0 - E7) from the kinematics of
extreme Population II stars (Binney, May \& Ostriker 1987b;
Sommer-Larsen \& Zhen 1990; van der Marel 1991; Amendt \& Cuddeford
1994).  From the dynamics of the precessing dusty disk of the S0 galaxy
NGC 4753 Steiman-Cameron \etal (1992) infer a rather round DM halo (E1). 
Buote \& Canizares (1996a, 1996b) used the shape of X-ray isophotes to infer
that the dark halos of NGC 1332 and NGC 720 are moderately flattened
(E5.5 and E6, respectively).  Cold dark matter galaxy formation
simulations which include gas dynamics tend to produce rather oblate DM
halos (Katz \& Gunn 1991; Udry \& Martinet 1994), with an intrinsic
flattening distribution peaked at $\qrho=c/a=0.5$ \pmt 0.15 (Dubinski
1994).  The current state of affairs is summarized in Fig.  1. 

\vspace*{-3mm}
\psfig{file=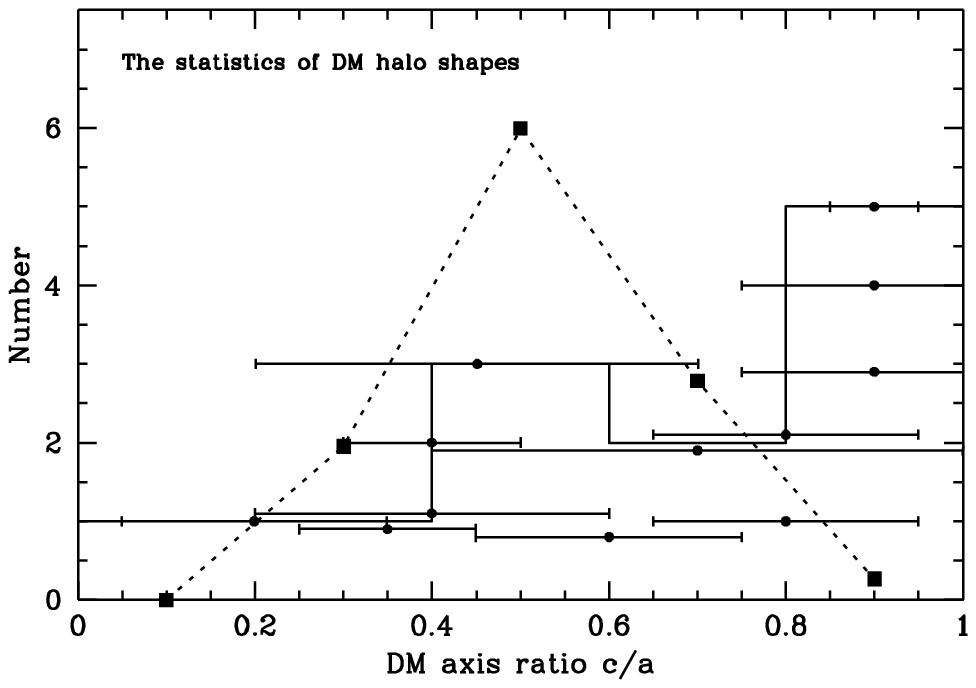,height=8cm,width=9cm}

\vspace*{-7.0cm}
\noindent
\parbox{3.7cm}{ Fig.  1. A histogram of the known DM halo shapes.  The
Dotted line and the filled squares represent the theoretical prediction. 
The points with error bars represent the individual galaxies.  Note the
discrepancy between the results from the warping-gas-layer method
(rightmost bin) and the other methods.}

\vspace*{1.0cm}

Van der Kruit (1981) pioneered the use of flaring measurements to
determine the mass of stellar disks, and found that the scale length of
the total matter (luminous plus dark) was similar to the scale length of
the light distribution and concluded that the mass-to-light ratio does
not vary significantly with radius.  This method could not be applied to
NGC 4244 since no reliable flaring information is obtained for the inner
parts of the galaxy.

In a previous paper, Paper I (Olling 1995), a method was developed to
determine the shape of the dark matter halo from the gaseous velocity
dispersion and the radial variation of the thickness of gas layers
(flaring).  This is accomplished by comparing the measured flaring with
that expected from a self-gravitating gaseous disk in the axisymmetric
potential due to the stellar disk and (flattened) DM halo.  That the
shape of the dark halo influences the width of the gas layer can be
easily understood.  Consider a round halo, with a certain density
distribution, then squeeze it along the vertical axis.  Consequently,
the densities as well as the exerted gravitational forces will increase,
resulting in a thinner HI disk and higher rotation speeds.  In order to
keep the same rotation curve, one has to deform the DM-halo in a very
specific way (Paper I) as a result of which the DM-halo densities (at
large distances) will be roughly inversely proportional to the
flattening \qrho.  Since the thickness of the gas layer beyond the
optical disk is proportional to $1/\sqrt{\rho_{DM}}$ (Paper I; cf.  Eqn. 
[1] with $\Kz{z} \approx 4 \pi G \int dz' \rho_{DM}(z') $), the halo
flattening $\propto$ (width of the gas layer)$^2$.

Below I apply the method to the galaxy NGC 4244 for which the basic
parameters were determined in Paper II (Olling 1996a).

\section{The Method}

Assuming that the gaseous velocity dispersion ($\sigma_{\rm gas}$) does
not vary with height above the plane ($z$), the gaseous density
distribution ($\rho_{\rm gas}(z)$) can be calculated from the equation of
hydrostatic equilibrium:

\vspace*{-5mm}
\begin{eqnarray}
\sigma_{\rm gas}^2 \; \frac{d \; \ln{\rho_{\rm gas}(z)} }{d \; z}  &=&
  -\Kz{(z)} \; \; ,
\end{eqnarray}
\vspace*{-2mm}

\noindent where the vertical force (\Kz{}) is calculated by integrating
over the density distribution of the galaxy $(\rho_{\rm tot}(R,z))$ :

\vspace*{-5mm}
\begin{eqnarray}
 \Kz{(R,z)} &=& G 
   \int_0^{\infty} r dr \rho(r,0) \int_{-\infty}^{\infty} dw \rho_{tot}(r,w)
   \int_{-\pi}^{\pi} \frac{d}{dz} 
   \frac{d\theta}{|\overline{\bf s} -\overline{\bf S}|} \; \; ,
\end{eqnarray}
\vspace*{-2mm}

\noindent with $\overline{\bf s}=\{r,w\}$ and $\overline{\bf
S}=\{R,z\}$.  Although more complicated than the more commonly used
local approximation (where the vertical force is calculated from the
local density distribution), this global approach has no problems in
those regions where the local approach fails: in the inner parts of the
galaxy where the rotation curve rises steeply and in the region where
the (stellar) density distribution is truncated (Paper I).  I
incorporate three components in the global mass model : 1) a double
exponential stellar disk with constant scale-height, 2) a non-singular
flattened isothermal DM-halo with core radius \Rc(\qrho) and central
density \rhoN(\qrho) (Paper I), and 3) a gaseous disk.  Three iterations
are required to determine $\rho_{\rm gas}(R,z)$ accurately.

The dependency of \Rc \ and $\rho_0$ upon \qrho \ is such that the
rotation curve of the flattened DM halo is practically indistinguishable
from its round equivalent (Paper I).  Of course, the true DM-halo
density distribution may be different (e.g., Navarro, Frenk, \& White
1996).  However, for roundish DM distributions the vertical force is
roughly proportional to the radial force ($\Kz = z/\sqrt{z^2 +R^2} \;
F_{\rm tot} \approx z/R \; F_{\rm R} \propto \frac{z}{R} \; V^2_{\rm
obs}$), which is the same for all disk-AnyRoundDarkHalo combinations
that reproduce the observed rotation curve: for a given rotation curve,
the width of the gas layer is independent of the radial distribution of
the DM.  The flattening of the DM halo introduces a $\sim
\sqrt{\qrho}$-dependence on the thickness of the gas layer, which might
be slightly different for various radial distributions. 

Comparing the thickness of the gas layer beyond the optical disk with
model flaring curves, calculated for a series of models with varying
halo flattening, then yields the halo shape.

\section{Results}

The almost edge-on, nearby Scd galaxy NGC 4244 was observed for about 14
hours with the VLA\VLAnote in B-, C-, and D- array configuration.  These
observations were used to determine the gaseous velocity dispersion, the
thickness of the gas layer, and the rotation curve (Fig.  2).  The
rotation curve of NGC declines from 5 optical scale-lengths to the last
measured point (at 8$h$) in Keplerian fashion (Olling 1996a).  While
compact fast rotating galaxies ($V_{max} \ge 180 \kms$) are known to
have declining rotation curves (Casertano \& van Gorkom 1991; Persic
Salucci, \& Stel 1996), NGC 4244 is the only low mass galaxy $(V_{max}
\approx 100 \ \kms)$ for which the rotation curve falls. 

\vspace*{-8mm}
\psfig{file=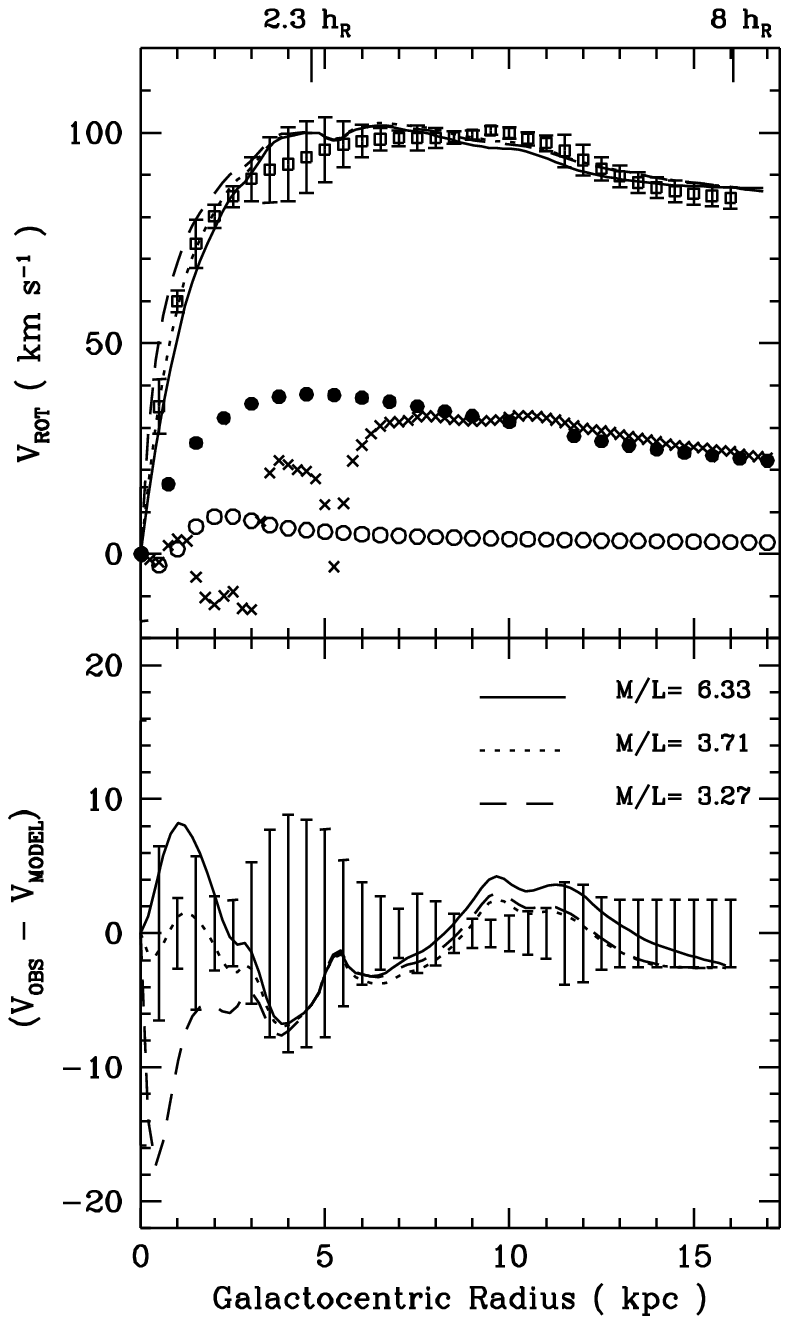,height=12cm,width=9cm}

\vspace*{-105mm} \noindent \parbox{4.8cm}{ Fig.  2.  The top panel
shows 3 disk-halo decompositions of the observed rotation curve (squares
with error bars) into components due to the gaseous disks (\HI,crosses;
H$_2$, open circles), a component due to the stellar disk (\MoverL{B}=1;
filled circles) and a DM component.  The individual stellar and DM
components are not shown here.  The best fit (\MoverL{} = 3.71, \Rc =
0.65 kpc, \rhoN = 233 m\Msun/pc$^3$, $\chi^2/DF$ = 0.864 with 32 degrees
of freedom, dotted line) as well as those models with a reduced $\chi^2$
value 1.0 larger are shown: \MoverL{} = 6.33, \Rc = 13, \rhoN = 1.43
(full line), and \MoverL{} = 3.27, \Rc = 0.28, \rhoN = 1300 (dashed
line).  The optical scale-length ($h_R$) and the truncation of the
stellar disk at $R_{\rm max}$ are also indicated. 
}

\vspace*{4mm}

I developed a new technique to determine simultaneously the thickness
and inclination of the \HI layer for galaxies at an inclination $\ge$ 60
degrees which uses about half the spectral line channels of the \HI data
set (Paper II).  The resulting flaring measurements for NGC 4244 are
presented in Fig.  3 for two cases.  An upper limit to the thickness of
the gas gayer is found by assuming a constant inclination of $84.5^o$
(open triangles), while incorporating the slight warp into the analysis
yields the best values for the flaring (filled triangles). 

Comparing the observations with the model flaring curves (drawn lines) I
conclude that the DM halo of NGC 4244 is highly flattened: \qrho = 0.2
\pmt 0.1, (Olling 1996b, hereafter referred to as Paper III).  The dark
halo of NGC 4244 is the flattest reported to date, furthermore it lies
at the extreme end of the theoretical predictions (Fig.  1).  Is it
possible that some systematic effect plays a role and that NGC 4244's DM
halo is less flattened? The most obvious candidate being an error in the
assumed inclination (Fig.  3, open triangles).  However, the warp is
very similar on both sides of the galaxy (Paper II), so that this is not
likely.  Another candidate is the presence of an extra galactic
radiation field (EgRF) which would ionize the \HI layer from above and
decrease the width of the {\em neutral} gas layer (Maloney 1993): if the
EgRF is as strong as the 2-$\sigma$ upper limit reported by Vogel \etal
(1995), we would infer a less flattened DM halo with \qrho = 0.5 \pmt
0.2 (Paper III).  Thirdly, non-thermal pressure gradients could be
important.  However, Bicay \& Helou (1990) find that cosmic rays are
closely related to sites of star formation so that cosmic ray pressure
(CRP) is not likely to be important beyond the optical disk.  If CRP
{\em is} important it would require an even {\em denser}, i.e.  {\em
flatter} DM halo.  Another possibility is that the gaseous velocity
dispersion tensor is anisotropic: if the vertical velocity dispersion is
smaller than the planar dispersion measured the DM halo would be rounder
than inferred above.  There is no observational evidence that such might
be the case.  Furthermore, because the Interstellar Medium (ISM) is
likely to be in the warm neutral phase due to the low pressure (Maloney
1993), the short collision times ($\leq 10^5$ year) preclude any
anisotropy in the velocity dispersion tensor. 

\vspace*{-20mm}
\psfig{file=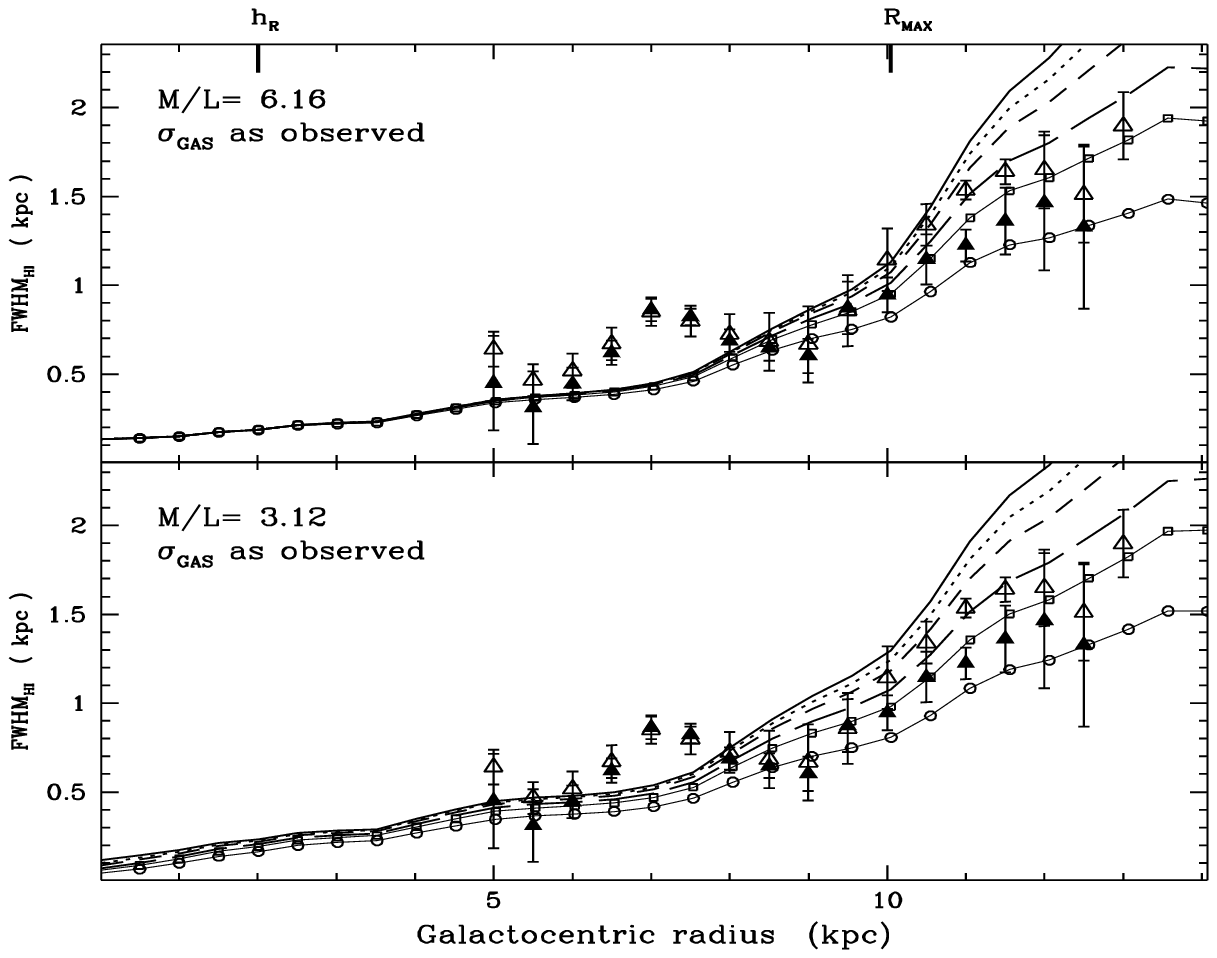,height=12cm,width=8.2cm}

\vspace*{-85mm} \noindent \parbox{4.8cm}{ Fig.  3.  
The measured gas layer widths (open and filled triangles for fixed
inclination and warp-included case, respectively).  The model curves
(drawn lines) correspond to different halo shape:\qrho=1.0, 0.7, 0.5,
0.3, 0.2, and 0.1 from top to bottom.  The measured gaseous velocity
dispersions (Paper II) were used in the model calculations.  Using both
inclination cases I find: \qrho = 0.2$_{-0.1}^{+0.3}$.  The
mass-to-light ratio of the stellar disk is not constrained by these
flaring measurements. 
}

\vspace*{15mm}

Taking systematic errors due to inclination and ionization effects into
account, we conclude that the DM halo of NGC 4244 is significantly
flattened, with $\qrho = 0.2_{-0.1}^{+0.3}$. 

\section{An Alternative Explanation ?}

Rich clusters of galaxies contain $\sim (10 \pm 5) h^{-1.5}$ \% (by
mass) hot X-ray emitting gas (e.g., Briel \etal 1992; Mushotzky \etal
1995), with $h$ the normalized Hubble constant ($h$ = H$_0$/(100 \kms
Mpc\rtp{-1})).  If these clusters are a ``fair sample'' of the universe
as a whole, then the average mass-to-light ratio of a cluster baryon
(\MoverL{Bar,cl}) lies in the range $(32 - 48) h^{-0.5}$.  Standard Big Bang
nucleosynthesis (BBN) models (e.g., Walker \etal 1991) limits
\MoverL{Bar,BBN} to the range $(11 - 35) h^{-1}$.  In the region where
the cluster and BBN estimates overlap ($h \geq 0.2$), $\MoverL{Bar} \sim
(35 \pm 6) h^{-1}$.  The dynamical mass-to-light ratios of individual
galaxies (Broeils 1992, 1995) range from 4 to 100 $h$, with 80\% of the
systems between 6 and 20 $h$.  It is thus possible that the dark halos
of individual galaxies consist mainly of non-baryonic dark matter, but a
100\% baryonic dark halo is also possible (Gott \etal 1974; Briel \etal
1992; Rubin 1993; Bahcall 1995; Sackett 1995).  It is not clear however
where, and in what form, these baryons reside in the galactic halos
since all plausible forms of baryonic dark matter seem to be excluded
(Hegyi \& Olive 1986). 

Pfenniger \etal (1994) reviewed cold, rotationally supported, molecular
hydrogen as a dark matter candidate.  In their model, small high density
molecular ``clumpuscules'' form the building blocks of a highly clumped
ISM.  Their model is best developed in the region beyond the optical
disk.  The fact that in many galaxies the shape of the rotation curve
due to the gas is similar to the observed rotation curve (Bosma 1981;
Carignan \etal 1990; Carignan \& Puche 1990) could then be explained if
only 3\% - 10\% of the gaseous surface density is in atomic form.  Such
might be expected in the context of clumpuscules hypothesis, or other
cold-gas dark matter models (e.g., Gerhard \& Silk 1995). 

Since the self-gravity of the gas layer beyond the optical disk strongly
affects the flaring (Paper I) I investigate whether the clumpuscules
hypothesis is consistent with NGC 4244's flaring curve.  Penny Sackett
(1995) kindly provided the disk-like surface density distribution
inferred from NGC 4244's rotation curve using a Keplerian as well as a
flat extrapolation beyond the last measured point ($\Sigma_{\rm tot,K}$
and $\Sigma_{\rm tot,F}$).  We find: $\Sigma_{\rm DM,K} \approx 86
\exp{(-R/5.2)}$, and $\Sigma_{\rm DM,F} \approx 55 \exp{(-R/11.4)}$
\MSpcsq.  For NGC 4244, the dark-to-\HI surface density is not constant:
$\Sigma_{DM}/\Sigma_{\protect\HI} \approx$ 10 exp(($R-10$)/(1.7 kpc)). 
Carignan \& Puche (1990) found a similar effect for NGC 7793.  This
contrasts Bosma's (1981) finding that the dark-to-\HI surface density
ratio is approximately constant with a value of 3-10\%. 

The thickness a dark gaseous disk can be calculated when a vertical
velocity dispersion is assumed.  I find that a dark disk has a thickness
equal to the \HI layer if the velocity dispersion of the dark disk is
1.1 (1.3) times larger than $\sigma_{\HI}$, for $\Sigma_{\rm DM,K}$
($\Sigma_{\rm DM,F}$).  With these dispersions the ``Keplerian dark
disk'' is close to being stable against radial instabilities ($Q = 0.8 -
1.2$ beyond the stellar disk) while the ``Flat dark disk'' is unstable
($Q = 0.8 - 0.3$).  Here I include a correction factor ($1+\frac{2
\pi}{R} ({\rm FWHM}_{\rm gas,z}/2.3) \approx 1.3$) due to the thickness
of the disk (e.g., Pfenniger \etal 1994).  Note that if the dark matter
is disk-like, it must have an anisotropic velocity dispersion tensor,
which may or may not be the case for clumpuscule-like dark matter. 

\section{Looking Ahead}

I have presented the results of a new method to determine the shape of
dark matter halos from sensitive \HI measurements and careful modeling. 
The first results exclude neither cold dark matter nor disk-like,
baryonic dark matter.  With current technology and ``reasonable''
observing times, the thickness of the \HI layer can be measured for
galaxies closer than $\sim$15 Mpc at inclinations $\ga 60^o$.  I
recently observed seven more systems (NGC 2366, 2403, 2903, 2841, 3521,
4236, and 5023) for which I will try to determine the DM halo shapes. 
Furthermore, the analysis of the flaring of the gas layers of the Milky
Way and M31 is in progress.  With this increased sample it will be
possible to gauge the significance of the highly flattened halo of NGC
4244 and, hopefully, put more stringent constraints on the nature of the
dark matter.

\acknowledgements

Most of the work presented here was part of my thesis project at
Columbia University.  I thank my advisor, Jacqueline van Gorkom, for
advise, guidance, lively discussions and friendship.  I thank Penny
Sackett for providing me with the disk-like surface density
distributions and Mike Merrifield for suggestions to improve this
contribution.  This work was supported in part through an NSF grant
(AST-90-23254 to J.  van Gorkom) to Columbia University and PPARC grant
GR/K58227.  And of course I like to thank the organizing committee of the
conference.

\end{document}